# Analysis of Quality of Service Performances of Connection Admission Control Mechanisms in OFDMA IEEE 802.16 Network using BMAP Queuing

**Abdelali EL BOUCHTI, Abdelkrim HAQIQ and Said EL KAFHALI**

**Computer, Networks, Mobility and Modeling laboratory**
**e- NGN research group, Africa and Middle East**
**FST, Hassan 1st University, Settat, Morocco**

**Abstract**

In this paper, we consider a single-cell IEEE 802.16 environment in which the base station allocates subchannels to the subscriber stations in its coverage area. The subchannels allocated to a subscriber station are shared by multiple connections at that subscriber station. To ensure the Quality of Service (QoS) performances, two Connection Admission Control (CAC) mechanisms, namely, threshold-based and queue-aware CAC mechanisms are considered at a subscriber station. A queuing analytical framework for these admission control mechanisms is presented considering Orthogonal Frequency Division Multiple Access (OFDMA) based transmission at the physical layer. Then, based on the queuing model, both the connection-level and the packet-level performances are studied and compared with their analogues in the case without CAC. The connection arrival is modeled by a Poisson process and the packet arrival for a connection by Batch Markov Arrival Process (BMAP). We determine analytically and numerically different QoS performance measures (connection blocking probability, average number of ongoing connections, average queue length, packet dropping probability, queue throughput and average packet delay).

## 1. Introduction

Worldwide Interoperability for Microwave Access (WiMAX) ([1], [2]) is becoming one of the hottest topics in the development of wireless technology. Researchers and developers are focusing on the development of WiMAX base station technology which is expected to provide services in 2008 around the world. In the 4th quantum of 2005, the IEEE 802.16e specification was launched to the market detailing the full specification of mobile WiMAX. In essence, the Quality of Service (QoS) for WiMAX is desperately required. WiMAX aims at providing low latency, low delay/jitter, low loss, adequate bandwidth service. In general, satisfactory QoS always requires a high operational cost. It is known that both Connection Admission Control (CAC) and packet scheduling co-operate to provide a high QoS and a low-cost service. To ensure further the QoS of high priority services, packet scheduling grants the channel for service according to their priorities so that un-serviced packets will line-up in the buffer.

Orthogonal Frequency Division Multiple Access (OFDMA) [5] is a promising wireless access technology for the next generation broad-band packet networks. With OFDMA [27], which is based on orthogonal frequency division multiplexing (OFDM), the wireless access performance can be substantially improved by transmitting data via multiple parallel channels, and also it is robust to inter-symbol interference and frequency-selective fading. OFDMA has been adopted as the physical layer transmission technology for WiMAX [26] based broadband wireless networks. Although the WiMAX standard [4] defines the physical layer specifications and the Medium Access Control (MAC) signaling mechanisms, the radio resource management methods such as those for CAC and dynamic bandwidth adaptation are left open. However, to guarantee QoS performances (e.g., call blocking rate, packet loss, and delay), efficient admission control is necessary in a WiMAX network [25] at both the subscriber and the base stations.

To analyze various connection admission control algorithms, analytical models based on continuous-time Markov chain (CTMC) ([3], [6], and [9]), were proposed in [19]. However, most of these models dealt only with call/connection-level performances [28] for the traditional voice-oriented cellular networks. In addition to the connection-level performances, packet-level performances also need to be considered for data-oriented packet-switched wireless networks such as WiMAX networks.

An earlier relevant work was reported by the authors in [23], [12], and [13]. They considered a similar model in OFDMA based-WiMAX but they modeled packet-level by Poisson process and MMPP process ([14], [16], and [17]) and they compared various QoS measures of CAC mechanisms. Since the introduction of batch Markovian arrival process (BMAP) by Lucantoni [21] many authors investigated queuing models with BMAP ([10], [15]). The reason is that BMAP enables more realistic and more accurate traffic modeling, since it can also capture dependency in traffic processes. Most of these works apply the standard matrix analytic-method pioneered by Neuts [18]. The incoming traffic has self similar and bursty nature also in wireless networks causing correlation in inter-arrival times, which influences the performance of the system. Our motivation for using BMAP is that it can model such traffic correlation. Hence applying BMAP in the queuing model enables the traffic correlation dependent performance evaluation of the system.

In this paper, we present two connection admission control mechanisms for a multi-channel and multi-user OFDMA network. The first mechanism is threshold-based, in which the concept of guard channel is used to limit the number of admitted connections to a certain threshold. The second mechanism, namely, queue-aware is based on the information on queue status and it also inherits the concept of fractional guard channel in which an arriving connection is admitted with certain connection acceptance probability. Specifically, the connection acceptance probability is determined based on the queue status (i.e., the number of packets in the queue). A queuing analytical model is developed based on a three- DTMC which captures the system dynamics in terms of the number of connections and queue status. We assume that the connection arrival and the packet arrival for a connection follow a Poisson process and a BMAP process respectively. Based on this model, various performance parameters such as connection blocking probability, average number of ongoing connections, average queue length, probability of packet dropping due to lack of buffer space, queue throughput, and average queuing delay are obtained. The numerical results reveal the comparative performance characteristics of the threshold-based and the queue-aware CAC and the without CAC algorithms in an OFDMA-based WiMAX network.

The remainder of this paper is organized as follows. Section 2 describes the system model including the objective of CAC policy. The formulation of the analytical model for connection admission control is presented in Section 3. In section 4 we determine analytically different performance parameters. Numerical results are stated in Section 5. Finally, section 6 concludes the paper.

## 2. Model description

### 2.1 System model

We consider a single cell in a IEEE 802.16/WiMAX network with a base station and multiple subscriber stations (Figure 1). Each subscriber station serves multiple connections. Admission control is used at each subscriber station to limit the number of ongoing connections through that subscriber station. At each subscriber station, traffic from all uplink connections are aggregated into a single queue [24]. The size of this queue is finite (i.e., $X$ packets) in which some packets will be dropped if the queue is full upon their arrivals. The OFDMA transmitter at the subscriber station retrieves the head of line packet(s) and transmits them to the base station. The base station may allocate different number of subchannels to different subscriber stations. For example, a subscriber station with higher priority could be allocated more number of subchannels.

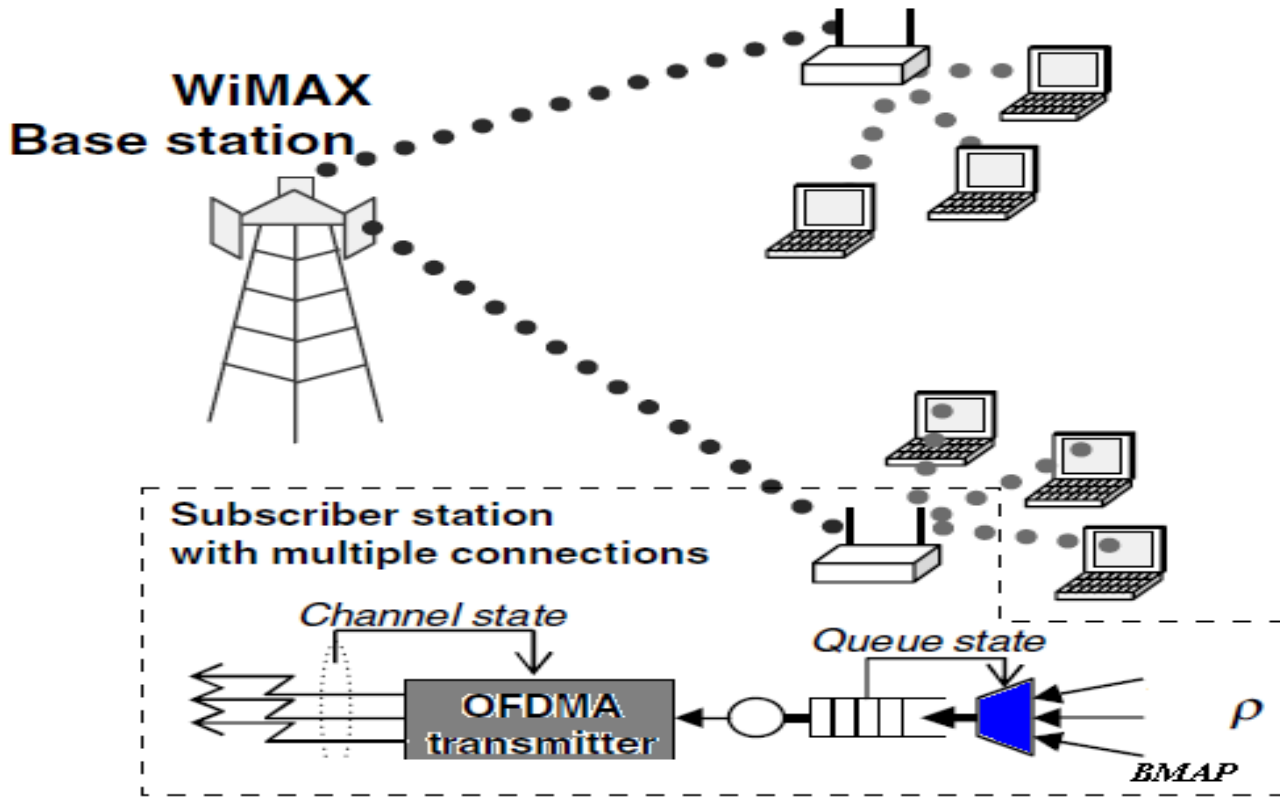

Figure 1: System model

### 2.1 CAC policy

The main objective of a CAC mechanism is to limit the number of ongoing connections/flows so that the QoS performances can be guaranteed for all the ongoing connections. Then, the admission control decision is made to accept or reject an incoming connection. To ensure the QoS performances of the ongoing connections, the following CAC mechanism for subscriber stations are proposed.

#### *2.1.1 Threshold-Based CAC mechanism*

In this case, a threshold C is used to limit the number of ongoing connections. When a new connection arrives, the CAC module checks whether the total number of connections including the incoming one is less than or equal to the threshold C. If it is true, then the new connection is accepted, otherwise it is rejected.

#### *2.1.2 Queue-Aware CAC mechanism*

This mechanism works based on connection acceptance probability $\alpha_x$ which is determined based on the queue status. Specifically, when a connection arrives, the CAC module accepts the connection with probability $\alpha_x$, where $x$ $(x \in \{0,1,...,X\})$ is the number of packets in the queue in the current time slot. Here, $X$ denotes the size of the queue of the subscriber station under consideration. Note that the value of the parameter $\alpha_x$ can be chosen based on the radio link level performance (e.g., packet delay, packet dropping probability) requirements.

## 3. Formulation of the Analytical Model

### 3.1 Formulation of the Queuing Model

An analytical model based on DTMC is presented to analyze the system performances at both the connection-level and at the packet-level for the connection admission ([7], [8], and [22]) mechanisms described before. We assume that packet arrival for a connection follows a BMAP process which is identical for all connections in the same queue. However, in the future we will consider non-homogenous Poisson process [29] as arrival process. The connection inter-arrival time and the duration of a connection are assumed to be exponentially distributed with average $1/\rho$ and $1/\mu$, respectively. In future, we will consider non exponential distributions using MRGP (Markov Re-Generative Process) [11] and/or phase-type expansions [29], [30].

The arrivals in the BMAP is directed by the irreducible continuous time Markov chain CTMC with a finite state space {0,1, …, $S$}. Sojourn time of the CTMC in the state $s$ has exponential distribution with parameter $\lambda_s$. After time expires, with probability $p_0(s,s')$ the chain jumps into the state $s'$ without generation of packets and with probability $p_k(s,s')$ the chain jumps into the state $s'$ and a batch consisting of k packets is generated, $k \geq 1$. The introduced probabilities satisfy conditions: $p_0(s,s)=0$, the sum of the probabilities of all outgoing transitions has to be equal to 1,

$$\sum_{k=1}^{\infty}\sum_{s'=0}^{S} p_k(s,s') + \sum_{\substack{s'=0 \\ s'\neq s}}^{S} p_0(s,s') = 1,\ 0 \leq s \leq S. \qquad (1)$$

The infinitesimal generator of BMAP is given as

$$Q_{BMAP} = \begin{pmatrix} D_0 & D_1 & D_2 & D_3 & \cdots \\ 0 & D_0 & D_1 & D_2 & \cdots \\ 0 & 0 & D_0 & D_1 & \cdots \\ 0 & 0 & 0 & D_0 & \cdots \\ \vdots & \vdots & \vdots & \vdots & \ddots \end{pmatrix} \qquad (2)$$

Where the matrices $D_k$ are given by

$$D_0 = [D_{ss'}],\ 0 \leq s \leq S,\ 0 \leq s' \leq S \qquad (3)$$

$$D_k = [D_{k,ss'}],\ 0 \leq s \leq S,\ 0 \leq s' \leq S,\ k \geq 0 \qquad (4)$$

Where:

$$D_{k,ss'} = \lambda_s p_k(s,s'),\ \ 0 \leq s \leq S,\ 0 \leq s' \leq S,\ \ k > 0. \qquad (5)$$

$$D_{ss} = -\lambda_s, \quad D_{ss'} = \lambda_s p(s,s'),\ s \neq s' \qquad (6)$$

Knowing the matrices $D_k$ the infinitesimal generator D can be defined as

$$D = \sum_{k=0}^{\infty} D_k. \qquad (7)$$

The steady-state probability vector $\pi_{BMAP}$ of the CTMC with generator D can be calculated as usual:

$$\pi_{BMAP}.D = 0, \quad \pi_{BMAP}.e = 1. \qquad (8)$$

Here and below $e$ is the column vector of appropriate dimension consisting of all 1's.
The mean steady-state arrival rate generated by the BMAP is:

$$\lambda_{BMAP} = \pi_{BMAP} \sum_{k=1}^{\infty} k D_k e. \qquad (9)$$

The state of the system is described by the process $Y_t = (S_t, X_t, C_t)$, where $S_t$ is the state (phase) of an irreducible continuous time Markov chain, $X_t$ is the number of packets in the aggregated queue and $C_t$ the number of ongoing connections at the end of every time slot $t$.

Thus, the state space of the system for both the CAC mechanisms is given by:

$$E = \{(s,x,c) / s \in \{1,....,S\},\ 0 \leq x \leq X,\ c \geq 0\}\ .$$

For the both CAC algorithms, the number of packet arrivals depends on the number of connections. However, for the queue-aware CAC algorithm, the number of packets in the queue affects the acceptance probability for a new connection. The state transition diagram is shown in figure 2.

Note that the probability that $n$ Poisson events with average rate $\rho$ occur during an interval T can be obtained as follows:

$$f_n(\rho) = \frac{e^{-\rho T}(\rho T)^n}{n!} \qquad (10)$$

This function is required to determine the probability of both connection and packet arrivals.

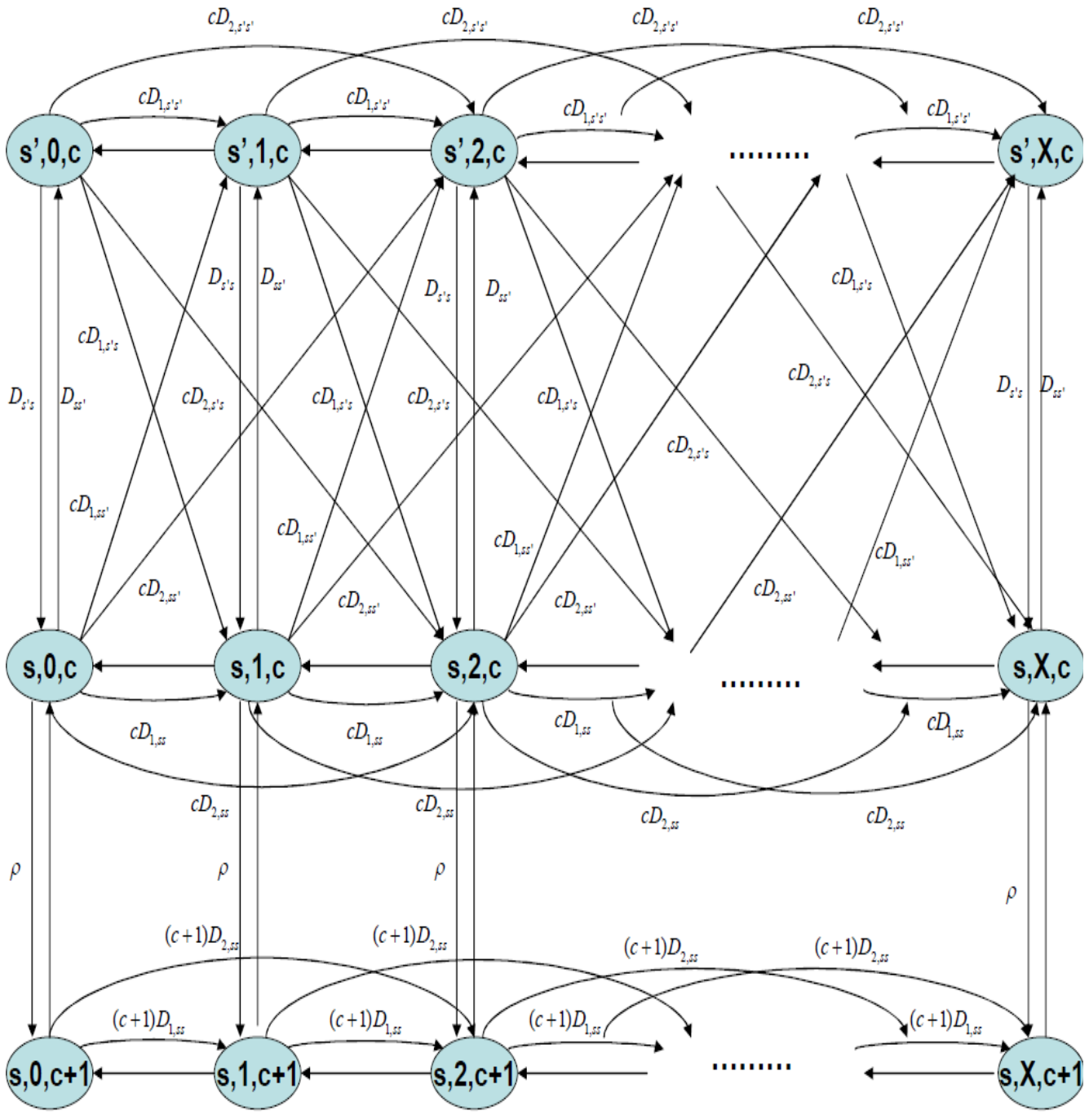

Figure 2: State transition diagram of discrete time Markov chain.

## 3.2 Threshold-Based CAC Algorithm

In this case, the transition matrix $Q$ for the number of connections in the system can be expressed as follows:

$$Q = \begin{bmatrix} q_{0,0} & q_{0,1} & & & & \\ q_{0,1} & q_{1,1} & q_{1,2} & & & \\ \ddots & \ddots & \ddots & & & \\ & & q_{C-2,C-1} & q_{C-1,C-1} & q_{C-1,C} \\ & & & q_{C-1,C} & q_{C,C} \end{bmatrix} \quad (11)$$

where each row indicates the number of ongoing connections. As the length of a frame $T$ is very small compared with connection arrival and departure rates, we assume that the maximum number of arriving and departing connections in a frame is one. Therefore, the elements of this matrix can be obtained as follows:

$$\begin{aligned} q_{c,c+1} &= f_1(\rho)\times(1-f_1(c\mu)),\ \ c=0,1,...,C\text{-}1 \\ q_{c,c-1} &= (1-f_1(\rho))\times f_1(c\mu),\ \ c=1,2,...,C \\ q_{c,c} &= f_1(\rho)\times f_1(c\mu)+(1-f_1(\rho))\times(1-f_1(c\mu)),\ \ c=0,1,...,C \end{aligned} \quad (12)$$

where $q_{c,c+1}$, $q_{c,c-1}$ and $q_{c,c}$ represent the cases that the number of ongoing connections increases by one, decreases by one, and does not change, respectively.

## 3.3 Queue-Aware CAC Algorithm

Because the admission of a connection in this case depends on the current number of packets in the queue, the transition matrix can be expressed based on the number of packets ($x$) in the queue as follows:

$$Q_x = \begin{bmatrix} q^{(x)}_{0,0} & q^{(x)}_{0,1} & & \\ q^{(x)}_{1,0} & q^{(x)}_{1,1} & q^{(x)}_{1,2} & \\ q^{(x)}_{2,0} & q^{(x)}_{2,1} & q^{(x)}_{2,2} & \\ & \ddots & \ddots & \ddots \end{bmatrix} \quad (13)$$

where:

$$\begin{aligned} q^{(x)}_{c,c+1} &= f_1(\alpha_x\rho)\times(1-f_1(c\mu)),\ \ c=0,1,... \\ q^{(x)}_{c,c-1} &= (1-f_1(\alpha_x\rho))\times f_1(c\mu),\ \ c=1,2,... \\ q^{(x)}_{c,c} &= f_1(\alpha_x\rho)\times f_1(c\mu)+(1-f_1(\alpha_x\rho))\times(1-f_1(c\mu)),\ \ c=0,1,... \end{aligned} \quad (14)$$

in which $\alpha_x$ is the connection acceptance probability when there are x packets in the queue.

## 3.4 Transition Matrix for the Queue

The transition matrix $P$ of the entire system can be expressed as in equation 15. The rows of matrix $P$ represent the number of packets ($x$) in the queue.

$$P = \begin{bmatrix} p_{0,0} & \cdots & p_{0,A} & & & & \\ \vdots & \vdots & \ddots & & \ddots & & \\ p_{R,0} & \cdots & p_{R,R} & \cdots & p_{R,R+A} & & \\ \ddots & \ddots & \ddots & \ddots & \ddots & & \ddots \\ & p_{x,x-R} & \cdots & p_{x,x} & \cdots & & p_{x,x+R} \\ & \ddots & \ddots & \ddots & \ddots & & \ddots \end{bmatrix} \quad (15)$$

Matrices $p_{x,x'}$ represent the changes in the number of packets in the queue (i.e., the number of packets in the queue changing from $x$ in the current frame to $x'$ in the next frame). We first establish matrices $v_{(s,x),(s,x')}$, where the diagonal elements of these matrices are given in equation 16. For $r \in \{0,1,2,...,L\}$ and $n \in \{0,1,2,...,(c\times A)\}$, $l = 1,2,...,L$, and $m=1,2,...,(c\times A)$. The non-diagonal elements of $v_{(s,x),(s,x')}$ are all zero.

$$\begin{aligned} \left[v_{(s,x);(s,x-l)}\right]_{c,c} &= \sum_{n-r=l}\sum_{k=1}^{\infty}\sum_{s'=0}^{S} f_n((c-1)D_{k,ss'})[R]_r \\ \left[v_{(s,x);(s,x+m)}\right]_{c,c} &= \sum_{r-n=m}\sum_{k=1}^{\infty}\sum_{s'=0}^{S} f_n((c-1)D_{k,ss'})[R]_r \\ \left[v_{(s,x);(s,x)}\right]_{c,c} &= \sum_{r=n}\sum_{k=1}^{\infty}\sum_{s'=0}^{S} f_n((c-1)D_{k,ss'})[R]_r \end{aligned} \quad (16)$$

Note that, matrix **R** [15] has size 1 × $R$ + 1, where $R$ indicates the maximum number of packets that can be transmitted in one frame. Here, $A$ is the maximum number of packets that can arrive from one connection in one frame and $L$ is the maximum number of packets that can be transmitted in one frame by all of the allocated sub channels allocated to that particular queue and it can be obtained from $L = \min\ (R, x)$. This is due to the fact that the maximum number of transmitted packets depends on the number of packets in the queue and the maximum possible number of transmissions in one frame. Note that, $\left[ \mathrm{v}_{(s,x);(s,x-l)} \right]_{c,c}$ , $\left[ \mathrm{v}_{(s,x);(s,x+m)} \right]_{c,c}$ and $\left[ \mathrm{v}_{(s,x);(s,x)} \right]_{c,c}$ represent the probability that the number of packets in the queue increases by $l$, decreases by $m$, and does not change, respectively, when there are $c$-$1$ ongoing connections. Here, $[\mathrm{v}]_{i,j}$ denotes the element at row $i$ and column $j$ of matrix v, and these elements are obtained based on the assumption that the packet arrivals for the ongoing connections are independent of each other.

Finally, we obtain the matrices $p_{x,x'}$ by combining both the connection-level and the queue-level transitions as follows:

$$p_{x,x'} = Q \mathrm{v}_{(s,x),(s,x')} \tag{17}$$

$$p_{x,x'} = Q_x \mathrm{v}_{(s,x),(s,x')} \tag{18}$$

for the cases of threshold-based (Equation 11) and queue-aware (Equation 13) CAC algorithms, respectively.

## 4. Performance Parameters

In this section, we determine the connection-level and the packet-level performance parameters (i.e., connection blocking probability, average number of ongoing connections in the system, and average queue length) for the both CAC mechanisms.

For the threshold-based CAC mechanism, all of the above performance parameters can be derived from the steady state probability vector of the system states $\pi$ , which is obtained by solving $\pi P = \pi$ and $\pi 1 = 1$ , where $1$ is a column matrix of ones. However, for the queue-aware CAC algorithm, the size of the matrix $Q_x$ needs to be truncated at $C_{tr}$ (i.e., the maximum number of ongoing connections at the subscriber station).

Also, the size of the matrix $P$ needs to be truncated at $X$ (i.e., the maximum number of packets in the queue) for the both mechanisms.

The steady-state probability, denoted by $\pi(s,x,c)$ for the state that there are $c$ connections and $x \in \{0,1,\ldots,X\}$ packets in the queue, can be extracted from matrix $\pi$ as follows

$$\pi(s,x,c) = [\pi]_{s \times x \times ((C+1)+c)}, \quad \mathrm{s} = 1,\ldots,S;\ c = 0,1,\ldots,C' \tag{19}$$

where $C' = C$ and $C' = C_{tr}$ for the threshold-based and the queue-aware CAC algorithms, respectively. Using these steady state probabilities, the various performance parameters can be obtained. Note that, the subscripts $tb$ and $qa$ are used to indicate the performance parameters for the threshold-based and the queue-aware CAC mechanisms, respectively.

### 4.1 Connection Blocking Probability

This performance parameter indicates that an arriving connection will be blocked due to the admission control decision. It indicates the accessibility of the wireless service, and for the threshold-based CAC mechanism. It can be obtained as follows:

$$p_{block}^{tb} = \sum_{s=1}^{S} \sum_{x=0}^{X} \pi(s,x,C). \tag{20}$$

The above probability refers to the probability that the system serves the maximum allowable number of ongoing connections.

The blocking probability for the queue-aware CAC mechanism is obtained from

$$p_{block}^{qa} = \sum_{s=1}^{S} \sum_{x=0}^{X} \sum_{c=1}^{C_{tr}} ((1-\alpha_x).\pi(s,x,C)). \tag{21}$$

in which the blocking probability is the sum of the probabilities of rejection for all possible number of packets in the queue.

### 4.2 Average Number of Ongoing Connections

It can be obtained as

$$N_c^{tb} = \sum_{s=1}^{S} \sum_{x=0}^{X} \sum_{c=0}^{C} c.\pi(s,x,c) \tag{22}$$

$$N_c^{qa} = \sum_{s=1}^{S} \sum_{x=0}^{X} \sum_{c=0}^{C_{tr}} c.\pi(s,x,c) \tag{23}$$

### 4.3 Average Queue Length Average

It is given by

$$N_x^{tb} = \sum_{s=1}^{S} \sum_{x=0}^{C} \sum_{x=0}^{X} x.\pi(s,x,c) \quad (24)$$

$$N_x^{qa} = \sum_{s=1}^{S} \sum_{x=0}^{C_{tr}} \sum_{x=0}^{X} x.\pi(s,x,c) \quad (25)$$

### 4.4 Packet Dropping Probability

This performance parameter refers to the probability that an incoming packet will be dropped due to the unavailability of buffer space. It can be derived from the average number of dropped packets per frame. Given that there are *x* packets in the queue and the number of packets in the queue increases by *m*, the number of dropped packets is $m-(X-x)$ for $m > X-x$ , and zero otherwise. The average number of dropped packets per frame is obtained as follows:

$$N_{drop} = \sum_{s=0}^{S}\sum_{c=1}^{C}\sum_{x=0}^{X}\sum_{m=X-x+1}^{A}\left(\sum_{l=1}^{C}[p_{x,x+m}]_{c,l}\right).(m-(X-x)).\pi(s,x,c) \quad (26)$$

where the term $\left(\sum_{l=1}^{C}[p_{x,x+m}]_{c,l}\right)$ indicates the total probability that the number of packets in the queue increases by *m* at every arrival phase. Note that, we consider probability $p_{x,x+m}$ rather than the probability of packet arrival as we have to consider the packet transmission in the same frame as well.

After calculating the average number of dropped packets per frame, we can obtain the probability that an incoming packet is dropped as follows:

$$p_{drop} = \frac{N_{drop}}{\overline{\lambda}} \quad (27)$$

where $\overline{\lambda}$ is the average number of packet arrivals per frame and it can be obtained from

$$\overline{\lambda} = \lambda_{BMAP} N_c \quad (28)$$

### 4.5 Queue throughput

It measures the number of packets transmitted in one frame and can be obtained from

$$\varphi = \lambda_{BMAP}(1-p_{drop}). \quad (29)$$

### 4.6 Average Packet Delay

It is defined as the number of frames that a packet waits in the queue since its arrival before it is transmitted. We use Little's law [22] to obtain average delay as follows:

$$D = \frac{N_x}{\varphi} \quad (30)$$

## 5. Numerical Results

In this section we present the numerical results of both CAC mechanisms. We use the Matlab software to solve numerically and to evaluate the various performance parameters.

### 5.1 Parameter Setting

We consider one queue (which corresponds to a particular subscriber station) for which five sub-channels are allocated and we assume that the average SNR is the same for all of these sub-channels. Each sub-channel has a bandwidth of 160 kHz. The length of a subframe for downlink transmission is one millisecond, and therefore, the transmission rate in one subchannel with rate ID = 0 (i.e., BPSK modulation and coding rate is 1/2) is 80 kbps. We assume that the maximum number of packets arriving in one frame for a connection is limited to 50.

For the threshold-based CAC mechanism, the value of the threshold C is varied according to the evaluation scenarios. For the queue-aware CAC mechanism, the value of the connection acceptance probability is determined as follows:

$$\alpha_{x=}\begin{cases} 1, & 0 \le x \prec B_{th} \\ 0, & B_{th} \le x \le X. \end{cases} \quad (30)$$

In the performance evaluation, we use $B_{th} = 100$.

For performance comparison, we also evaluate the queuing performance in the absence of CAC mechanism. For the case without CAC, we truncate the maximum number of ongoing connections at 70. The average duration of a connection is set to twenty minutes for all the evaluation scenarios. The queue size is 300 packets. The parameters are set as follows: The connection arrival rate is 0.9 connections per minute the batch packets of size 30 (i.e., *k*=30). Average SNR on each sub-channel is 5 db.
For clarity, the all numerical parameters are summarized in table 1.

Table 1: Summary of numerical parameters.

| | |
|---|---|
| $X$ | 300 |
| $A$ | 50 |
| $C_{tr}$ | 70 |
| $B_{th}$ | 100 |
| $\rho$ | 0.9 |
| $k$ | 30 |
| $\mu$ | 20 |
| Average SNR | 5 dB |

Note that, we vary some of these parameters depending on the evaluation scenarios whereas the others remain fixed.

## 5.2 Performance of CAC policy

We first examine the impact of connection arrival rate on connection-level performances. Variations in average number of ongoing connections and connection blocking probability with connection arrival rate are shown in Figures 3 and 4. As expected, when the connection arrival rate increases, the number of ongoing connections and connection blocking probability increase.

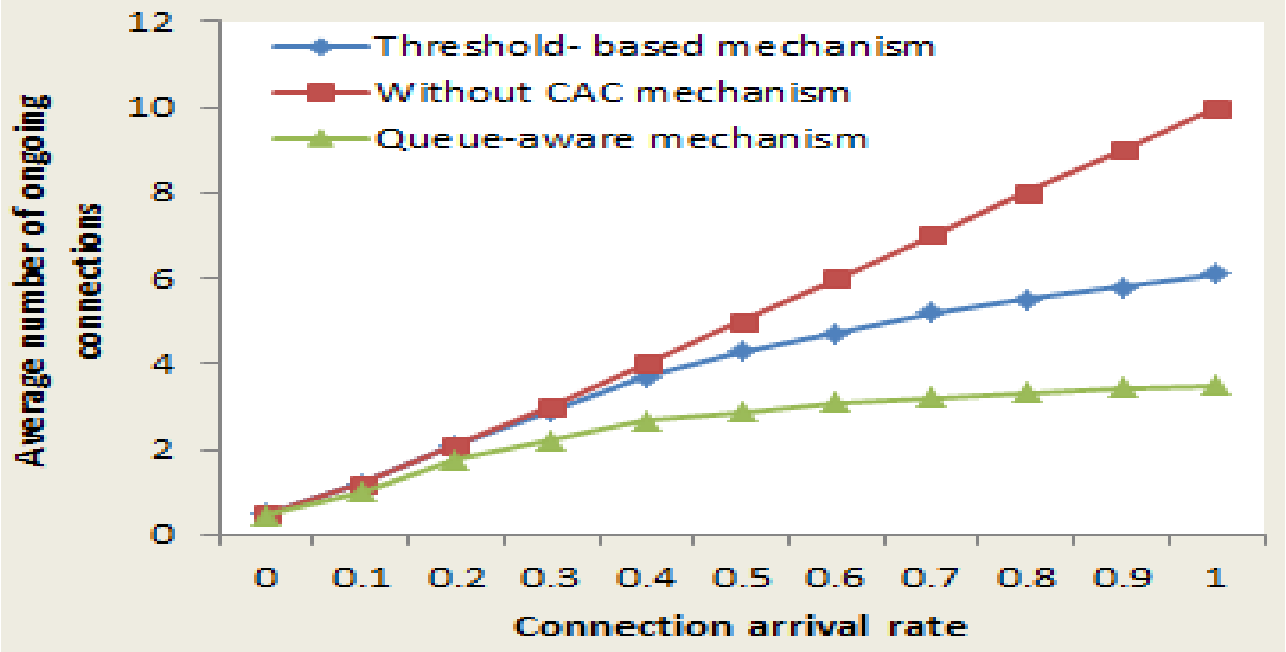

Figure 3: Average number of ongoing connections under different connection arrival rates.

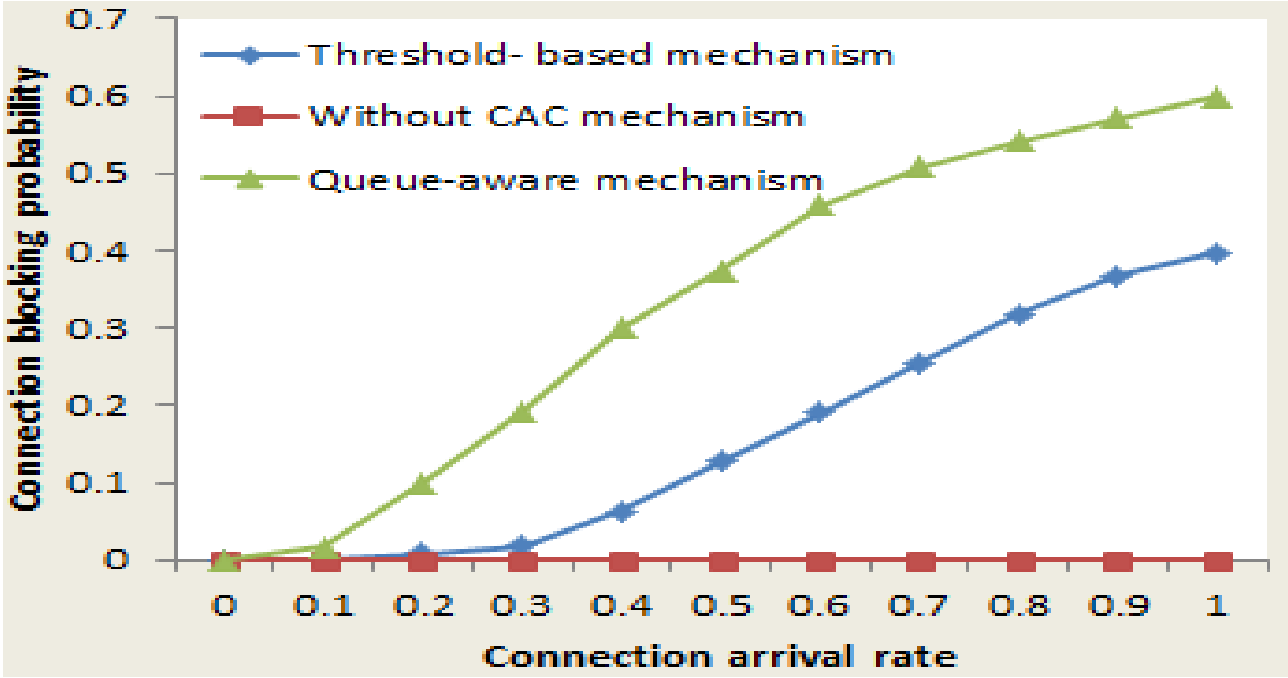

Figure 4: Connection blocking under different connection arrival rates.

The packet-level performances under different connection arrival rates are shown in Figures 5 through 8 for average number of packets in the queue, packet dropping probability, queue throughput, and average queuing delay, respectively. These performance parameters are significantly impacted by the connection arrival rate. Because the both CAC mechanisms limit the number of ongoing connections, packet-level performances can be maintained at the target level. In this case, both CAC mechanisms result in better packet-level performances compared with those without CAC mechanism.

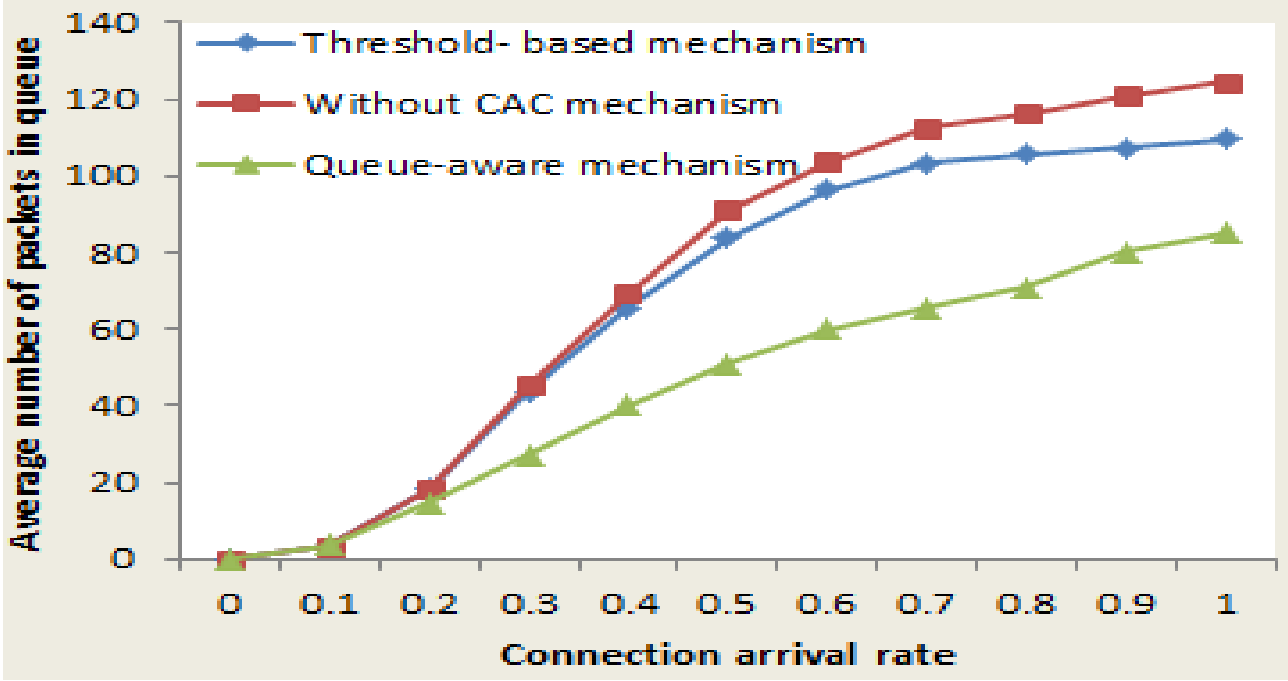

Figure 5: Average number of packets in queue under different connection rates.

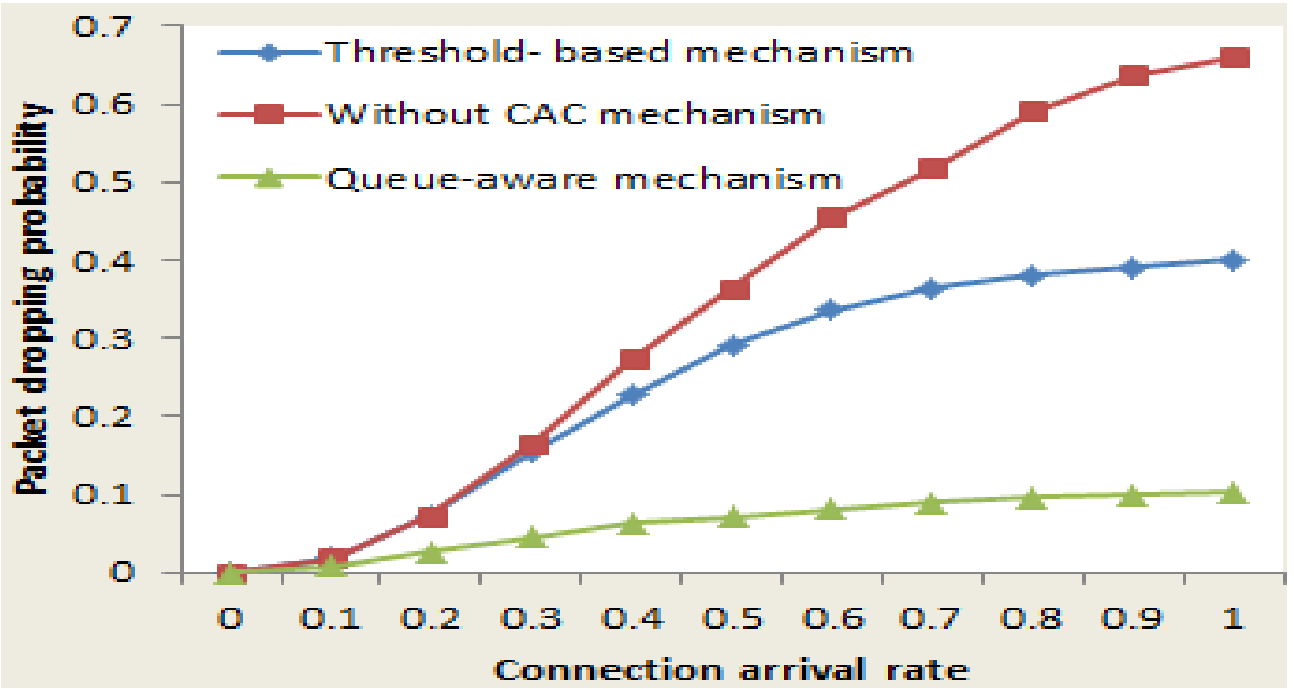

Figure 6: Packet dropping under different connection arrival rates.

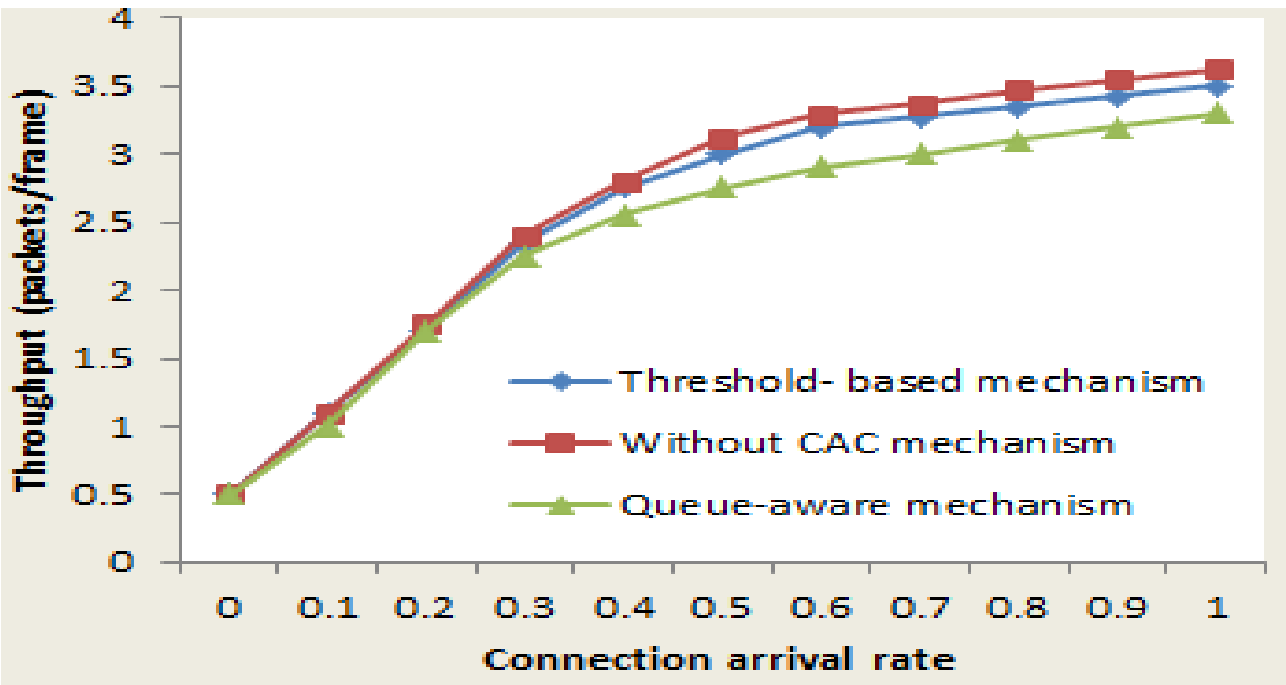

Figure 7: Queuing throughput under different connection arrival rates.

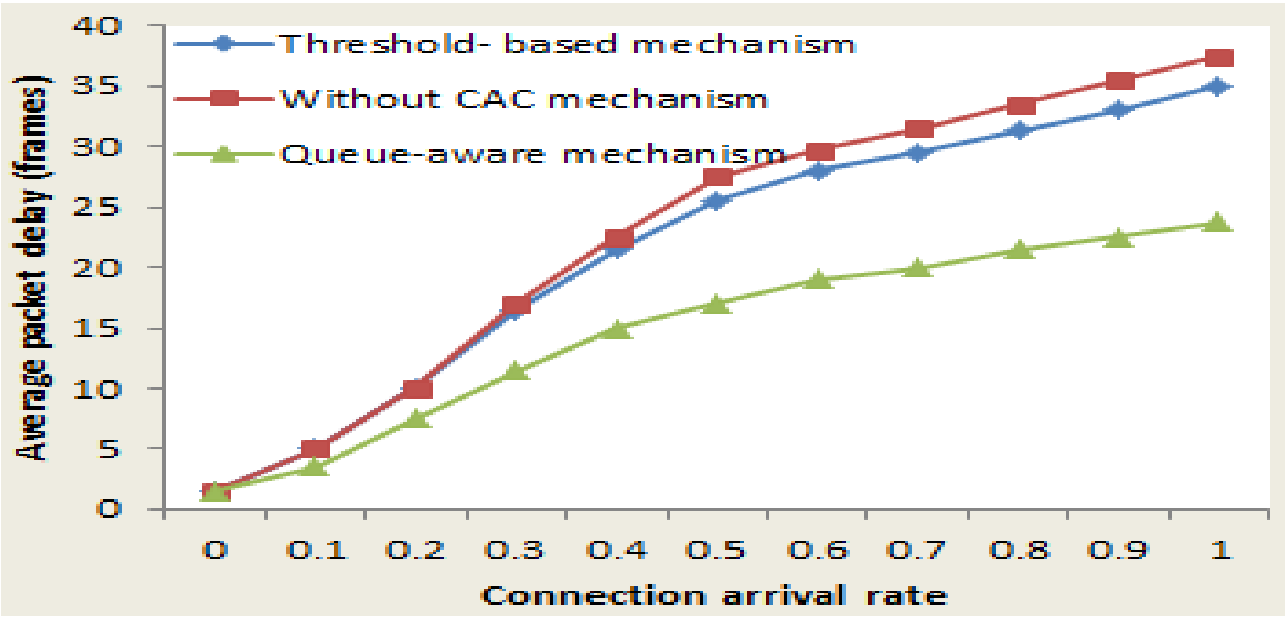

Figure 8: Average packet delay under different connection arrival rates.

Variations in packet dropping probability and average packet delay with channel quality are shown in Figures 9 and 10. As expected, the packet-level performances become better when channel quality becomes better.

Also, we observe that the connection-level performances for the threshold-based CAC mechanism and those without CAC mechanism are not impacted by the channel quality when this later becomes better (the connection blocking probability remains constant when the channel quality varies), connection blocking probability decreases significantly for the queue-aware CAC mechanism when the channel quality becomes better (Figure. 11).

Based on these observations, we can conclude that the queue-aware CAC can adapt the admission control decision based on the queue status which is desirable for a system with high traffic fluctuations.

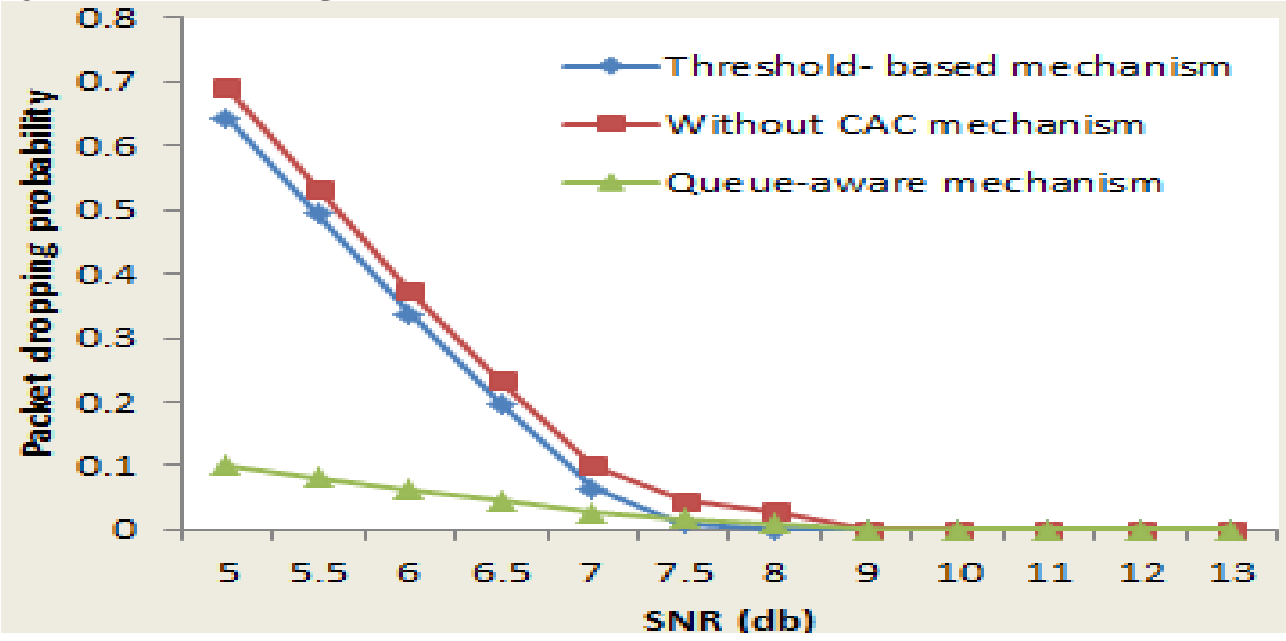

Figure 9: Packet dropping probability under different channel qualities.

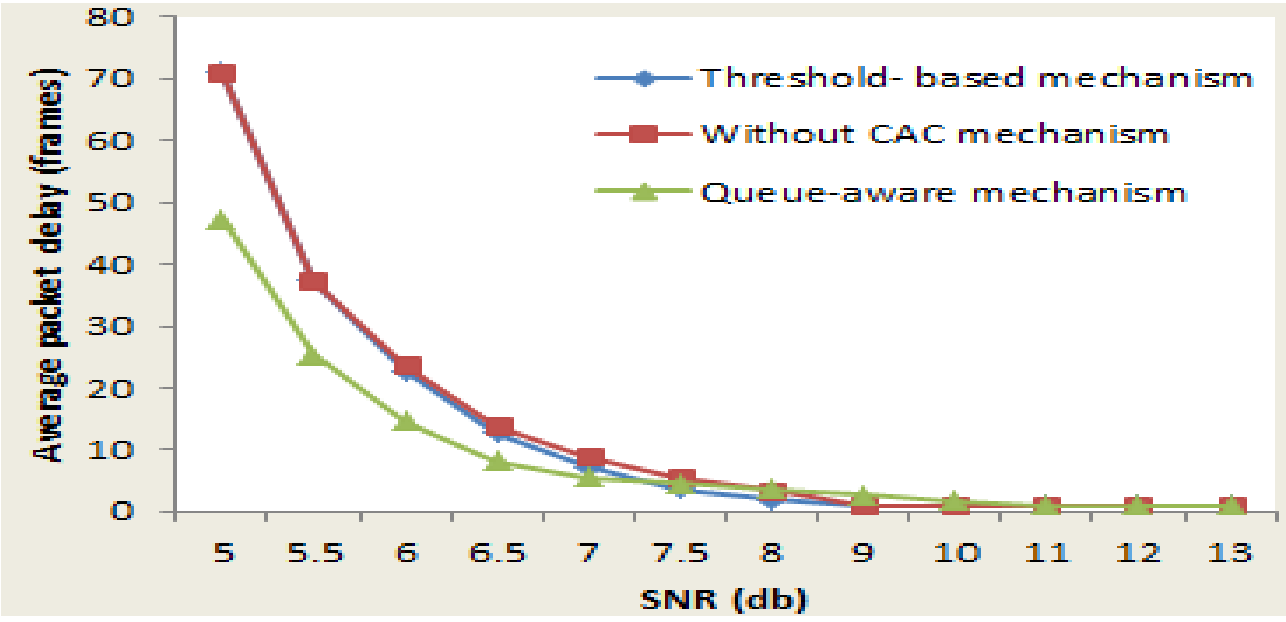

Figure 10: Average packet delay under different channel qualities.

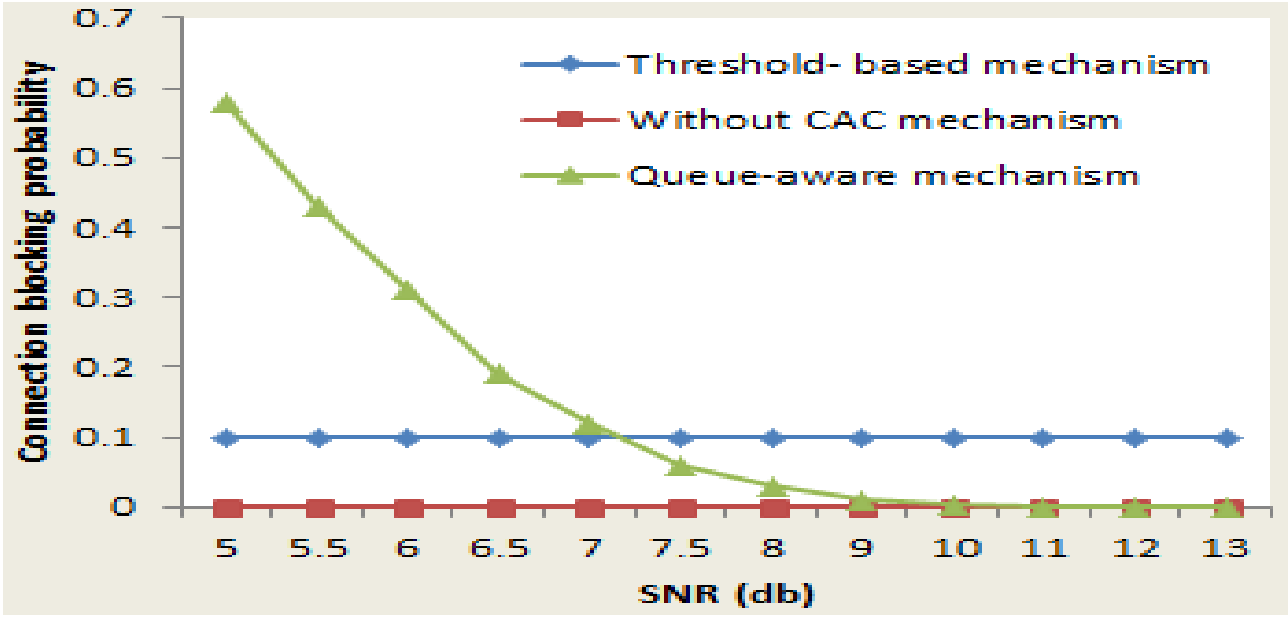

Figure 11: Connection blocking probability under different channel qualities.

## 6. Conclusion

In this paper, we have addressed the problem of queuing theoretic performance modeling and analysis of OFDMA transmission under admission control. We have considered a IEEE 802.16/WiMAX system model in which a base station serves multiple subscriber stations and each of the subscriber stations is allocated with a certain number of subchannels by the base station. There are multiple ongoing connections at each subscriber station.

We have presented two connection admission control mechanisms for a multi-channel and multi-user OFDMA network, namely, queue-aware mechanism and threshold-based mechanism. While the threshold-based CAC mechanism simply fixes the number of ongoing connections, the queue-aware CAC mechanism considers the number of packets in the queue for the admission control decision of a new connection. The connection-level and packet-level performances of these CAC mechanisms have been studied based on the queuing model.

The connection-level and packet-level performances of the both CAC mechanisms have been studied based on the queuing model. The connection arrival is modeled by a Poisson process and the packet arrival for a connection by a BMAP process. We have determined analytically and numerically different performance parameters, such as connection blocking probability, average number of ongoing connections, average queue length, packet dropping probability, queue throughput, and average packet delay.

Numerical results show that, the performance parameters of connection-level and packet-level are significantly impacted by the connection-level rate, the both CAC mechanisms results in better packet-level performances compared with those without CAC mechanism. The packet-level performances become better when channel quality becomes better. On the other hand, the connection-level performances for the threshold-based CAC mechanism and those without CAC mechanism are not impacted by the channel quality when this later becomes better. Then, the queue-aware CAC can adapt the admission control decision based on the queue status which is desirable for a system with high traffic fluctuations.

All the results showed in this paper remain in correlation with those presented in [23], [12] and [13] even if we change here the arrival packet Poisson process by an MMPP process or by BMAP process, which is more realistic.